\documentclass[reprint, superscriptaddress, twocolumn, amsmath, amssymb, aps, prb]{revtex4-2}
\usepackage{graphicx}
\usepackage{dcolumn}
\usepackage{bm}
\usepackage{placeins}
\usepackage{appendix}
\usepackage{upgreek}

\usepackage{verbatim}
\usepackage[usenames,dvipsnames]{xcolor}
 \usepackage{amsmath}
 \usepackage{amsfonts}
 \usepackage{amssymb}
 \usepackage[colorlinks=true,citecolor=blue,linkcolor=red]{hyperref}

 \usepackage{bbold}     
 \usepackage[makeroom]{cancel}  
 \usepackage{multirow}    
 \usepackage[normalem]{ulem}        
 \usepackage{array}
 \usepackage{pdfpages}
\makeatletter
 \AtBeginDocument{\let\LS@rot\@undefined}
 \makeatother

\begin{document}


\title{Enhanced superconductivity in PbTe-In hybrids}

\author{Zuhan Geng}
\email{equal contribution}
\affiliation{State Key Laboratory of Low Dimensional Quantum Physics, Department of Physics, Tsinghua University, Beijing 100084, China}

\author{Fangting Chen}
\email{equal contribution}
\affiliation{State Key Laboratory of Low Dimensional Quantum Physics, Department of Physics, Tsinghua University, Beijing 100084, China}

\author{Yichun Gao}
\email{equal contribution}
\affiliation{State Key Laboratory of Low Dimensional Quantum Physics, Department of Physics, Tsinghua University, Beijing 100084, China}

\author{Lining Yang}
\email{equal contribution}
\affiliation{State Key Laboratory of Low Dimensional Quantum Physics, Department of Physics, Tsinghua University, Beijing 100084, China}

\author{Yuhao Wang}
\affiliation{State Key Laboratory of Low Dimensional Quantum Physics, Department of Physics, Tsinghua University, Beijing 100084, China}

\author{Shuai Yang}
\affiliation{State Key Laboratory of Low Dimensional Quantum Physics, Department of Physics, Tsinghua University, Beijing 100084, China}

\author{Shan Zhang}
\affiliation{State Key Laboratory of Low Dimensional Quantum Physics, Department of Physics, Tsinghua University, Beijing 100084, China}

\author{Zonglin Li}
\affiliation{State Key Laboratory of Low Dimensional Quantum Physics, Department of Physics, Tsinghua University, Beijing 100084, China}

\author{Wenyu Song}
\affiliation{State Key Laboratory of Low Dimensional Quantum Physics, Department of Physics, Tsinghua University, Beijing 100084, China}

\author{Jiaye Xu}
\affiliation{State Key Laboratory of Low Dimensional Quantum Physics, Department of Physics, Tsinghua University, Beijing 100084, China}

\author{Zehao Yu}
\affiliation{State Key Laboratory of Low Dimensional Quantum Physics, Department of Physics, Tsinghua University, Beijing 100084, China}

\author{Ruidong Li}
\affiliation{State Key Laboratory of Low Dimensional Quantum Physics, Department of Physics, Tsinghua University, Beijing 100084, China}

\author{Zhaoyu Wang}
\affiliation{State Key Laboratory of Low Dimensional Quantum Physics, Department of Physics, Tsinghua University, Beijing 100084, China}

\author{Xiao Feng}
\affiliation{State Key Laboratory of Low Dimensional Quantum Physics, Department of Physics, Tsinghua University, Beijing 100084, China}
\affiliation{Beijing Academy of Quantum Information Sciences, Beijing 100193, China}
\affiliation{Frontier Science Center for Quantum Information, Beijing 100084, China}
\affiliation{Hefei National Laboratory, Hefei 230088, China}

\author{Tiantian Wang}
\affiliation{Beijing Academy of Quantum Information Sciences, Beijing 100193, China}
\affiliation{Hefei National Laboratory, Hefei 230088, China}

\author{Yunyi Zang}
\affiliation{Beijing Academy of Quantum Information Sciences, Beijing 100193, China}
\affiliation{Hefei National Laboratory, Hefei 230088, China}

\author{Lin Li}
\affiliation{Beijing Academy of Quantum Information Sciences, Beijing 100193, China}

\author{Runan Shang}
\affiliation{Beijing Academy of Quantum Information Sciences, Beijing 100193, China}
\affiliation{Hefei National Laboratory, Hefei 230088, China}

\author{Qi-Kun Xue}
\affiliation{State Key Laboratory of Low Dimensional Quantum Physics, Department of Physics, Tsinghua University, Beijing 100084, China}
\affiliation{Beijing Academy of Quantum Information Sciences, Beijing 100193, China}
\affiliation{Frontier Science Center for Quantum Information, Beijing 100084, China}
\affiliation{Hefei National Laboratory, Hefei 230088, China}
\affiliation{Southern University of Science and Technology, Shenzhen 518055, China}

\author{Ke He}
\email{kehe@tsinghua.edu.cn}
\affiliation{State Key Laboratory of Low Dimensional Quantum Physics, Department of Physics, Tsinghua University, Beijing 100084, China}
\affiliation{Beijing Academy of Quantum Information Sciences, Beijing 100193, China}
\affiliation{Frontier Science Center for Quantum Information, Beijing 100084, China}
\affiliation{Hefei National Laboratory, Hefei 230088, China}

\author{Hao Zhang}
\email{hzquantum@mail.tsinghua.edu.cn}
\affiliation{State Key Laboratory of Low Dimensional Quantum Physics, Department of Physics, Tsinghua University, Beijing 100084, China}
\affiliation{Beijing Academy of Quantum Information Sciences, Beijing 100193, China}
\affiliation{Frontier Science Center for Quantum Information, Beijing 100084, China}


\begin{abstract}

We report the realization of epitaxial indium thin films on PbTe nanowires. The film is continuous and forms an atomically sharp interface with PbTe. Tunneling devices reveal a hard superconducting gap. The gap size, 1.08-1.18 meV, is twice as large as bulk indium’s ($\sim$ 0.5 meV), due to the presence of PbTe. A similar enhancement is also observed in the critical temperature of In on a PbTe substrate. Subgap peaks appear at finite magnetic fields. The effective $g$-factor (15-45) is notably enhanced compared to bare PbTe wires ($<$ 10) due to the presence of In, differing from Al-hybrids. Josephson devices exhibit gate-tunable supercurrents. The PbTe-In hybrid enhances the properties of both, the superconductivity of In and g-factors of PbTe, and thus may enable exotic phases of matter such as topological superconductivity. 

\end{abstract}

\maketitle  

\section{Introduction}

Superconductor and semiconductor devices are the backbone of solid-state quantum computing. The combination of the two materials inherits properties of both, such as superconducting correlation, gate-tunable carrier density, and low-dimensional geometry. These ingredients can be engineered in quantum devices, such as hybrid nanowires, for the exploration of Majorana zero modes \cite{Lutchyn2010, Oreg2010, Mourik, Deng2016, Gul2018,  Song2022, WangZhaoyu,Delft_Kitaev, MS_2023, NextSteps} and hybrid qubits \cite{2015_PRL_gatemon,DiCarlo_gatemon,2019_PRX_Scalay, 2021_Devoret_Science, Huo_gatemon,2023_NP_Andreev,Ramon_perspective}. The superconductor that has been mostly studied is aluminum, which can form a pristine interface with semiconductors \cite{Krogstrup2015}. Al can induce a hard superconducting gap \cite{Chang2015, PanCPL}, crucial for high-quality devices \cite{Takei2013}. Thin Al films exhibit a larger gap \cite{1965_Al_gap, 1971_Al, 2008_Al_gap} and a higher critical (in-plane) magnetic field ($B$). Despite these merits, Al-based nanowires are, however, facing great challenges in Majorana search. The gap size (below 0.4 meV for thin films) is still small and fragile to disorder \cite{Patrick_Lee_disorder_2012, Prada2012, Loss2013ZBP, Liu2017, Loss2018ABS, GoodBadUgly, DasSarma_estimate, DasSarma2021Disorder, Tudor2021Disorder}. The spin-orbit interaction is weak, and the $g$-factor of 2 is small. The metalization  effect in Al-hybrids suppresses the spin-orbit interaction and g-factors in nanowires \cite{Loss_metalization}. These disadvantages motivate the search for alternative superconductors with superior properties.

Niobium-based superconductors have been extensively studied \cite{Gul2017,Zhang2017Ballistic, CPH_Shadow}. Although the gap size is larger (0.5-0.9 meV) \cite{Gul2017,Zhang2017Ballistic, CPH_Shadow, Gul2018},  the issue of soft gap \cite{Takei2013} arises upon increasing $B$, due to vortex formation \cite{Jouri2019}. Epitaxial Sn induces a gap of 0.5-0.7 meV in InSb or InAs nanowires \cite{InSb-Sn, InAs-Sn}. The presence of non-superconducting phase ($\alpha$-Sn) imposes challenges for future applications. Pb is a promising superconductor with a large induced gap (1.0-1.3 meV) \cite{InAs-Pb, Yichun}. Promising Majorana signatures have yet to be observed. 

Indium is a common superconductor whose properties are not attractive at first sight. Its critical temperature $T_{\text{c}}$ of 3.4 K corresponds to a gap $\Delta  = 1.764 k_{\text{B}}T_{\text{c}}\sim$ 0.5 meV ($k_{\text{B}}$ is the Boltzmann constant). The gap in InAs-In hybrids is even smaller (0.45 meV) \cite{CPH_In}. The formation of disconnected grains on InAs \cite{CPH_In} poses a serious issue for quantum devices. The granular morphology hinders the realization of NS tunneling devices which require a long, thin, and continuous In film (N stands for normal metal and S for superconductor). The low melting temperature ($\sim$ 157 $^{\circ}$C) is another obstacle for device fabrication.

In this study, we tackle these challenges by realizing continuous thin In films, epitaxially grown on PbTe. We choose PbTe as the semiconductor due to its advantage in disorder mitigation compared to InAs or InSb, see Refs. \cite{CaoZhanPbTe, Jiangyuying, Erik_PbTe_SAG, PbTe_AB, Fabrizio_PbTe, Zitong, Wenyu_QPC, Yichun, Yuhao, Ruidong, Vlad_PbTe, Wenyu_Disorder} for recent progress. The In film forms an atomically sharp interface with PbTe, resulting in a hard superconducting gap. The gap size, $\Delta \sim$ 1.08-1.18 meV, is enhanced by more than a factor of two compared to InAs-In hybrids  \cite{CPH_In} or the bulk In. The $T_c$ of In thin film on a PbTe substrate ($\sim$ 6 K) is also significantly enhanced compared to the case of other substrates ($\sim$ 3.8 K). The enhanced superconductivity of In on PbTe is the key observation of this work.  Subgap states in tunneling conductance can be observed at finite $B$'s. The effective g-factors are also enhanced due to the presence of In, beneficial for Majorana realization \cite{Loss_metalization}.  

\section{Material and device characterization}

\begin{figure*}[htb]
\includegraphics[width=\textwidth]{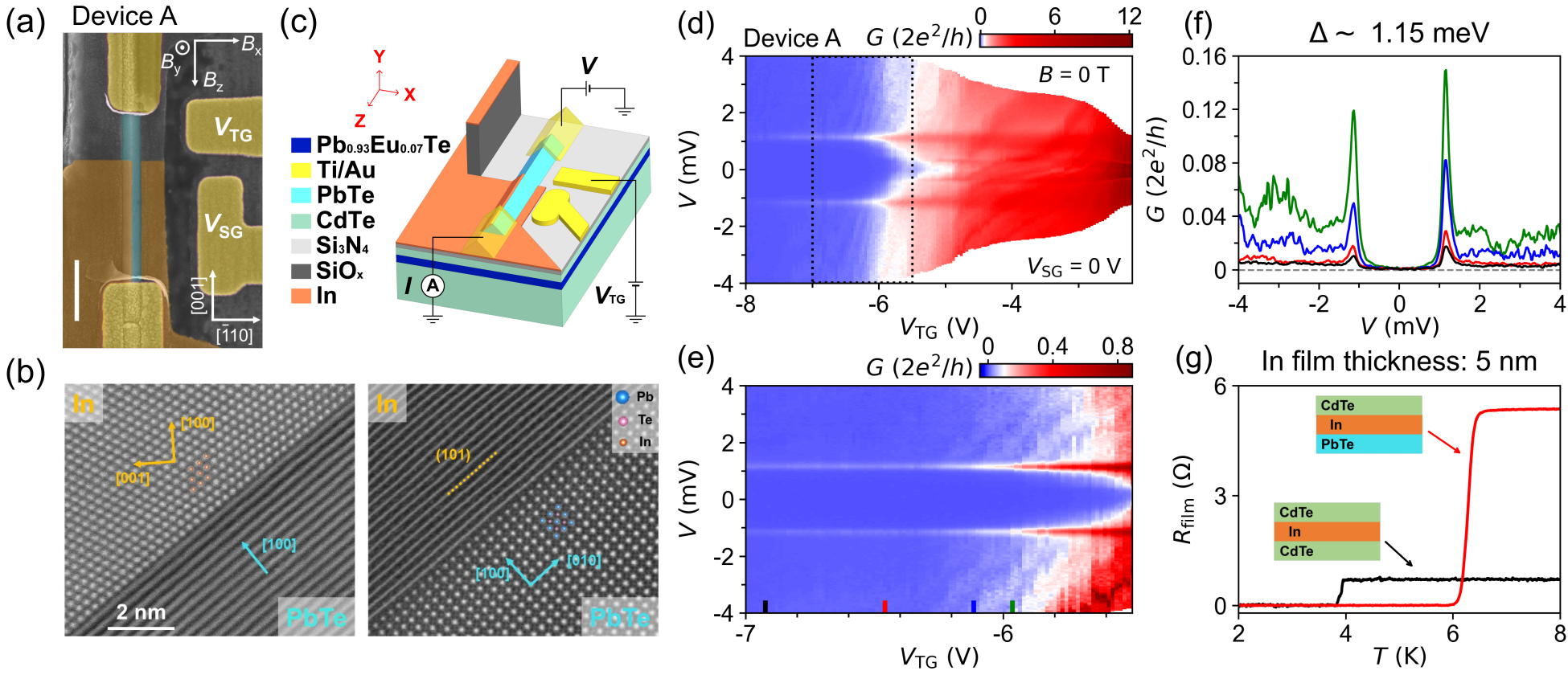}
\centering
\caption{PbTe-In interface, hard gap, and $T_c$ enhancement. (a) False-colored SEM of device A. Scale bar, 500 nm.  (b) STEM of PbTe-In interface.  (c) Device schematic (CdTe capping not drawn for clarity). (d) $G$ vs $V$ and $V_{\text{TG}}$ of device A.  (e) Fine scan in the tunneling regime (the dashed box in (d)). (f) Line cuts from (e), see the color bars. (g) $T_c$ measurement of In films on PbTe (red) and CdTe (black) substrates.  }
\label{fig1}
\end{figure*}

Figure 1(a) shows a scanning electron micrograph (SEM) of an NS device. The material growth followed Ref. \cite{Wenyu_Disorder} with minor modifications. The PbTe nanowires were grown selectively on a Pb$_{0.93}$Eu$_{0.07}$Te/CdTe(110) substrate, followed by the in situ shadow-wall deposition of In (thickness 7-10 nm). The chip was then capped by 10-nm-thick CdTe. The sample stage was cooled using liquid nitrogen throughout the In growth and capping. The capping prevents In oxidation and aggregation, keeping the film morphology homogeneous. For device fabrication, hot-plate baking was avoided due to the low melting point of In. Electron-beam resist thus underwent vacuum pumping at room temperature. Contacts and side gates were then fabricated by evaporating Ti/Au, during which the sample stage was cooled using liquid nitrogen. Additional information can be found in the Supplemental Material \cite{SM}. 

Figure 1(b) is the scanning transmission electron microscopy (STEM) of the interface of a PbTe-In nanowire. To have atomic resolution, the images were taken along the [010] zone axis of In (left) and [001] zone axis of PbTe (right), respectively. The interface is atomically sharp with no interdiffusion, crucial for high-quality devices. For additional STEM results, see Fig. S1 \cite{SM}.

Figure 1(c) is a schematic of an NS device  (CdTe capping not drawn for clarity). Standard two-terminal measurement was carried out in a dilution fridge (base temperature $<$ 50 mK). Figure 1(d) shows the differential conductance, $G\equiv dI/dV$, as a function of $V$ and the tunnel gate voltage $V_{\text{TG}}$. $I$ is the current and $V$ is the bias drop across the device. $V_{\text{SG}}$ = 0 throughout the measurement.  We find no obvious formation of quantum dots for this device from the open regime to pinched-off. The features at $V = \pm 1.15$ mV indicate the gap edges.  Figure 1(e) is a fine measurement in the tunneling regime, with several line cuts shown in Fig. 1(f). A hard gap with sharp coherence peaks is revealed. The gap size, $\Delta \sim$ 1.15 meV, is more than twice of the bulk In's (0.5 meV) and the InAs-In hybrids (0.45 meV) \cite{CPH_In}.  This twofold enhancement is unexpected, as the $T_{\text{c}}$ for In films with similar thickness is only slightly enhanced \cite{1961_Indium,1967_Indium, 1971_Indium, 2009_Indium}.

To gain insights on the unexpectedly-large-enhancement, we grew In thin films (thickness 5 nm) on flat CdTe and PbTe substrates, see the inset of Fig. 1(g). The films are continuous and capped by CdTe, see Fig. S2 for images \cite{SM}. Figure 1(g) shows the $T_c$ measurement of the two films using four-terminal method. The black curve denotes the case of CdTe substrate, revealing a $T_c$ of 3.8 K. This value is consistent with literature on In films with similar thickness \cite{1961_Indium,1967_Indium, 1971_Indium, 2009_Indium}, and is slightly enhanced compared to bulk In (3.4 K). The red curve is the case of PbTe substrate. The $T_c$ of  $\sim$ 6 K is surprisingly larger compared to the regular case (black curve).

\section{Thickness dependence of In films}

\begin{figure}[htb]
\includegraphics[width=\columnwidth]{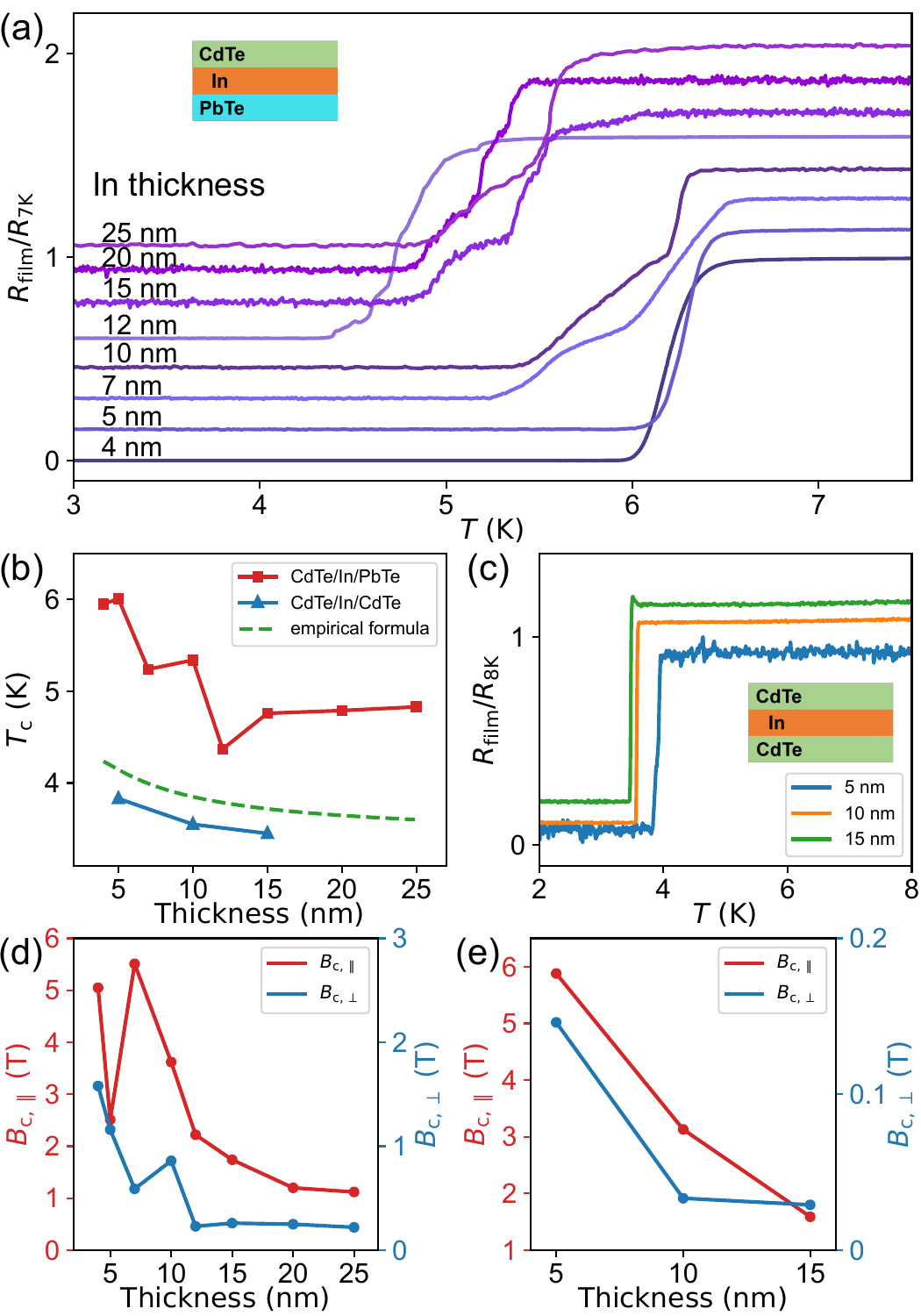}
\centering
\caption{Thickness and magnetic field dependence of In films. (a) In on PbTe with various film thickness. Vertical offset, 0.15. (b) $T_c$ extracted from (a) (the red scatters) and (c) (the blue scatters). The green dashed line is the formula from literature.  (c) In on CdTe. Vertical offset, 0.1. (d-e) In-plane ($B_{c, \parallel}$) and out-of-plane ($B_{c, \perp}$) critical fields of In films on PbTe (d) and CdTe (e) substrates. }
\label{fig2}
\end{figure}

In Figure 2, we study In films with varying thicknesses on PbTe and CdTe substrates. Figure 2(a) shows the $T_c$ measurements for In films on PbTe with different film thicknesses. The film resistance is normalized to $R_{\text{film}}/R_{7\text{K}}$ (where $R_{7\text{K}}$ is the resistance at 7 K).  The curves are vertically offset for clarity. The extracted $T_c$'s are shown as red dots in Fig. 2(b).  The blue dots represent $T_c$'s of In on CdTe, extracted from Fig. 2(c). The green dashed line corresponds to the empirical formula, $T_c (\text{K}) = 3.405 + 5.2/d + 7.5/d^2$, taken from Toxen \cite{1961_Indium}, where $d$ is the film thickness in nm. While $T_c$ generally decreases with increasing thickness for both substrates, the $T_c$ for In on PbTe is significantly higher than both the empirical formula and the $T_c$ for In on CdTe within the explored thickness range. The pronounced difference in $T_c$ (PbTe-In vs CdTe-In) and gap size (PbTe-In vs InAs-In \cite{CPH_In}) indicates a substantial enhancement in superconductivity for In due to the presence of PbTe. The possible underlying mechanism might be due to the large dielectric constant of PbTe and charge transfer between the two materials. 

Charge transfer at heterostructure interfaces has been proposed as a key mechanism for enhanced superconductivity in various material systems such as copper oxides \cite{Tc_1} and single unit-cell FeSe films \cite{Tc_2, Tc_3, Tc_4}. Additionally, interfacial electron-phonon coupling may play a role, either through softened phonon modes in reduced dimensions \cite{Tc_5} or via interactions between superconducting electrons and substrate phonons \cite{Tc_4, Tc_6}. Further theoretical investigations could provide deeper insights into the enhanced superconductivity observed in In/PbTe systems.

Figures 2(d-e) show the critical magnetic fields of the films, measured in both in-plane ($B_{c, \parallel}$) and out-of-plane ($B_{c, \perp}$) directions. Additional data can be found in Fig. S2. Thinner films are able to sustain higher critical fields, suggesting a reduced orbital effect. Note that the maximum field applied was is 6 T due to hardware limitations, which means the 6 T data point in Fig. 2(e) is underestimated. 

\section{Magnetic field dependence of the induced gap}

\begin{figure}[htb]
\includegraphics[width=\columnwidth]{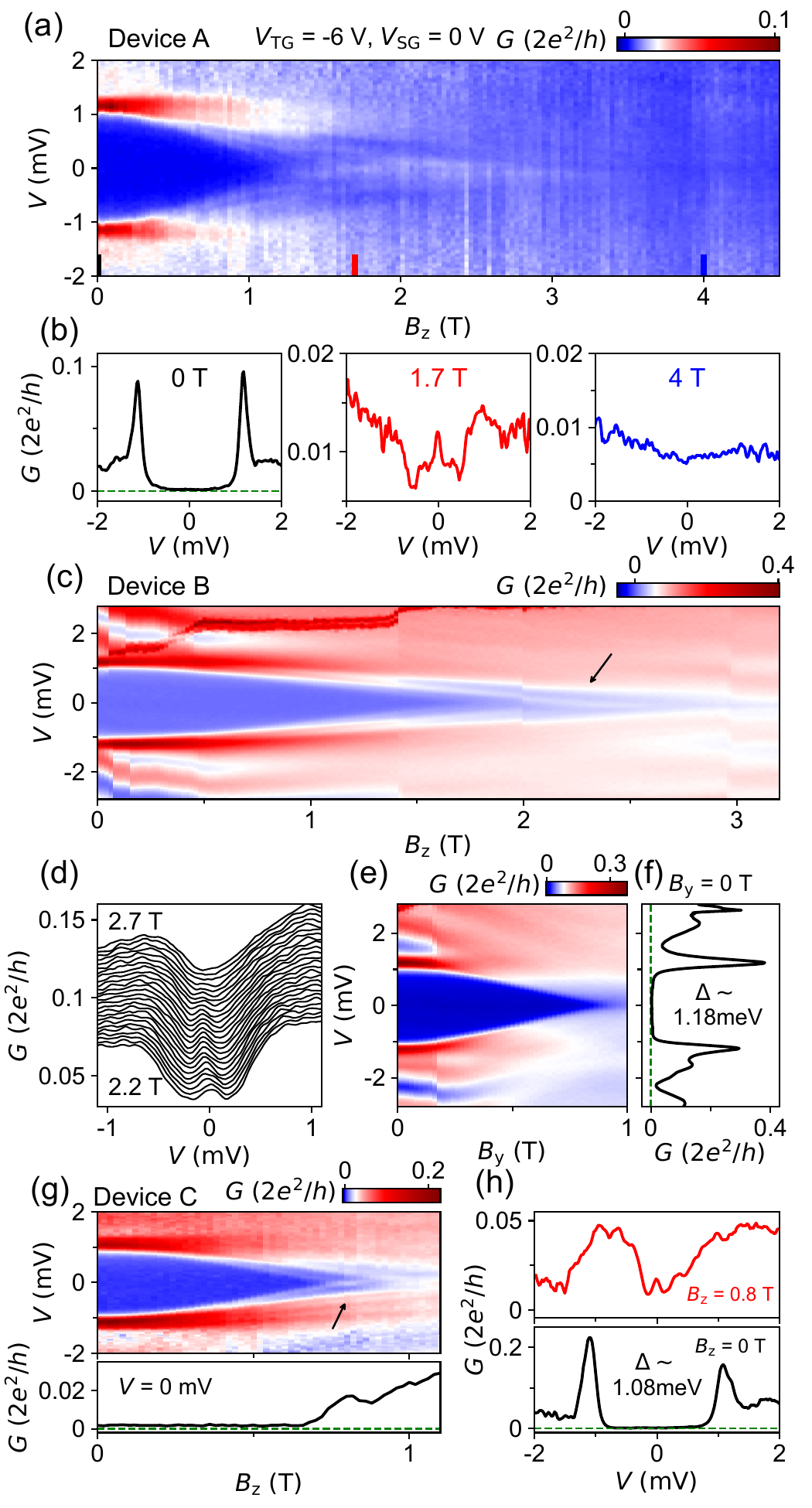}
\centering
\caption{Subgap states and g-factors. (a) $G$ vs $V$ and $B_z$ of device A. (b) Line cuts from (a). (c) $B$ scan of device B. (d) Line cuts from (c). Vertical offset, 0.003$\times 2e^2/h$ between neighboring curves. (e) $B_y$ scan of the gap. $V_{\text{TG}}$ = -0.6 V for (c) and (e). (f) Line cut from (e) at 0 T. (g) $B$ scan of device C. Lower panel, zero-bias line cut. $V_{\text{TG}}$ = 4.7 V. (h) Line cuts from (g). }
\label{fig2}
\end{figure}

In Fig. 3(a), we study the PbTe-In nanowire (device A) and scan $B$ along the $z$-axis (roughly aligned with the nanowire). The axis orientation is sketched in Fig. 1(a). Two Andreev levels detach from the gap edges and merge to zero  at $\sim$ 1.2 T, accompanied by gap softening. To extract the effective $g$-factor of Andreev levels, we focus on the $B$ region with a linear dispersion relation, see Fig. S3 for its linear fitting.  A $g$-factor of $\sim$ 28.5 can be extracted based on the slopes of Andreev levels \cite{Michiel2018, CPH2018effectiveg}. In our previous studies on bare PbTe wires, g-factors are highly anisotropic, and can be estimated based on the sizes of ballistic quantized plateaus at finite $B$'s \cite{Yuhao}. For wires with the same crystal orientation as the devices in this work (i.e. PbTe on a CdTe(110) substrate) and the same $B$ direction (along the wire), a smaller $g$-factor ($<$ 10) was extracted for 5 PbTe devices  \cite{Yuhao}. Here, the $g$-factor is significantly enhanced due to the presence of In, beneficial for Majoranas as it can lower the critical $B$. In contrast, the g-factors in InAs-Al and InSb-Al devices are usually suppressed in the tunneling regime \cite{Deng2016, Michiel2018, WangZhaoyu} compared to the bare III-V wires due to metallization \cite{Loss_metalization}, unless if the device was tuned into the weak coupling regime where a soft and smaller gap emerges \cite{LutchynSchrodinger, Flensberg_SP, DasSarma_SP}.

The suggap peaks merge toward zero and form a small and non-robust zero-bias peak, see Fig. 3(b) (middle panel) for a line cut at 1.7 T. The peak likely originates from an Andreev bound state (ABS) \cite{Silvano2014} or disorder \cite{Patrick_Lee_disorder_2012,GoodBadUgly}. At higher $B$'s, e.g. 4 T in Fig. 3(b), the gap is fully closed. For $B$ scans along $x$ and $y$ axes, see Fig. S4 \cite{SM}.

Figure 3(c) shows the $B$ scan of a second NS device, see Fig. S5 for its SEM and gate scan \cite{SM}. A smaller $g$-factor of $\sim$ 15 is estimated. The subgap states thus cross zero at higher $B$'s (2.2 - 2.6 T), see Fig. 3(d) for waterfall plot. The gap survives for $B >$ 3 T. For the $y$-axis, the gap closes at $\sim$ 1 T (Fig. 3(e)), possibly due to the orbital effect of In film. The gap is $\sim$ 1.18 meV at 0 T (Fig. 3(f)). 

Figure 3(g) shows the result of a third device. $V_{\text{TG}}$ = 4.7 V. The subgap states cross zero at $\sim$ 0.8 T, see Fig. 3(h) for a line cut. Consequently, the $g$-factor (along the nanowire) is $\sim$ 45, more than a factor of three compared to bare PbTe wires. For additional data  of device C, see Fig. S6 \cite{SM}. 

\section{Josephson devices}

We next explore Josephson devices (SS geometry). Figure 4(a) shows an SEM of device D. The junction width is $\sim$ 270 nm, and was formed by shadow-wall deposition \cite{Zitong}. The In film is continuous and $\sim$ 1.3 $\upmu$m long on each side of the junction. Figure 4(b) shows the $I$-$V$ curve over a large $I$-range. The red dashed line is a linear fit for $V >$ 2.2 mV. The slope of the fit gives an estimation of the normal state resistance $R_{\text{n}} \sim$ 2.68 k$\Omega$.  The intercept of the fit (extrapolation) on $I$-axis gives an estimation of the excess current $I_{\text{excess}} \sim$ 0.53 $\upmu$A. We then calculate $eI_{\text{excess}}R_{\text{n}}/\Delta \sim$ 1.2, corresponding to a junction transparency of $\sim$ 80$\%$ \cite{Flensberg_1988}.  

\begin{figure}[b]
\includegraphics[width=\columnwidth]{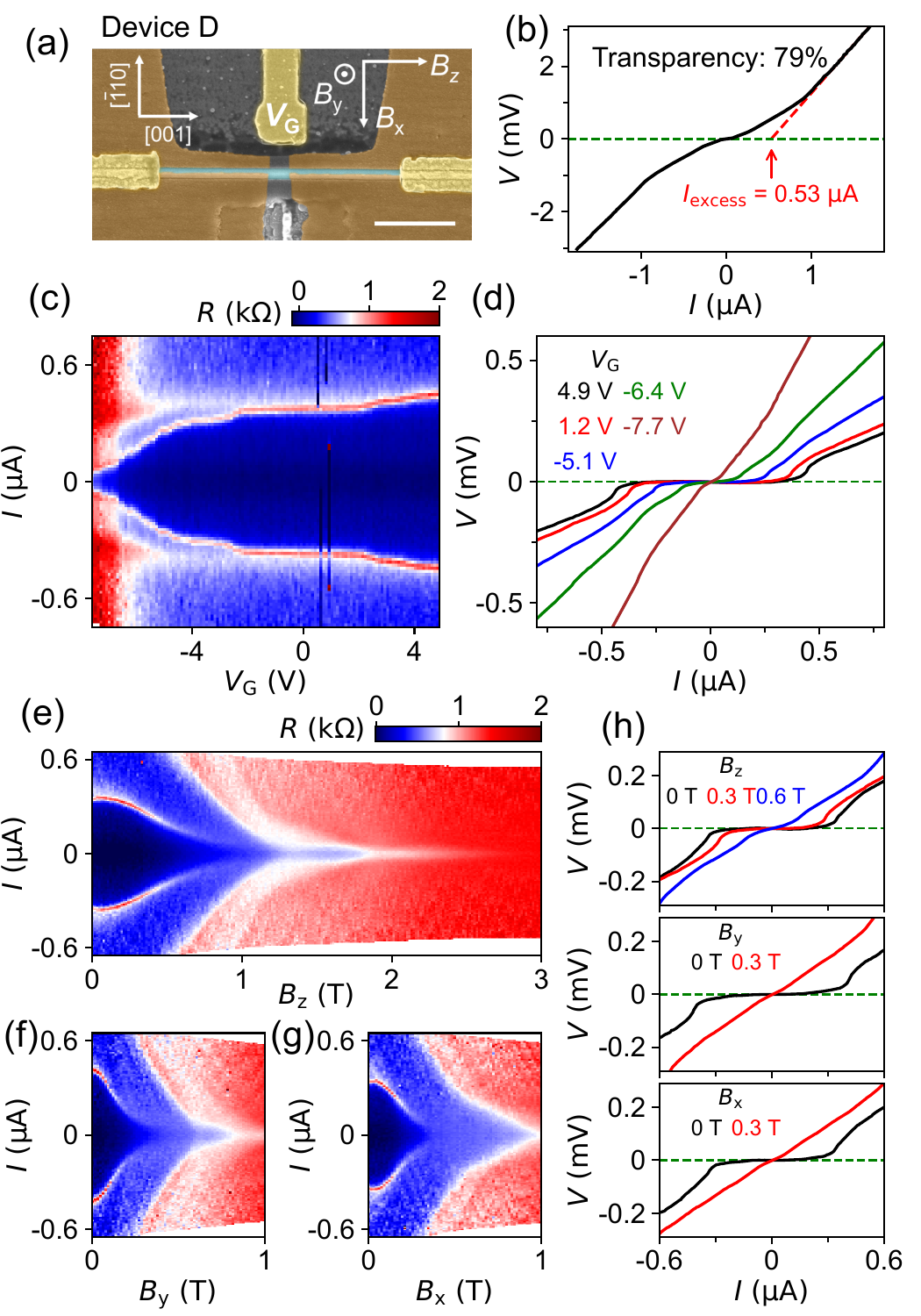}
\centering
\caption{Josephson device. (a) False-colored SEM of device D. Scale bar, 1 $\upmu$m. (b) $I$-$V$ curve at $V_{\text{G}}$ = -6.8 V. The red dashed line is a linear fit for $V>2\Delta/e$. (c) $R$ vs $I$ and $V_{\text{G}}$. $B$ = 0 T. (d) Line cuts from (c), shown as $I$-$V$ curves. (e-g) $B$ dependence of the supercurrent for $x$, $y$ and $z$ axes. $V_{\text{G}}$ = 3 V. (h) Line cuts from (e-g).   }
\label{fig3}
\end{figure}

Figure 4(c) shows the gate dependence of supercurrent. For a better visibility, differential resistance $R \equiv dV/dI$ is plotted, by numerically differentiating the $I$-$V$ curves. For line cuts in $I$-$V$ format, see Fig. 4(d). The switching current can be gate-tuned from 0 to 400 nA. The corresponding Josephson energy, $E_J = \hbar I_c/2e \sim$ 380 $\upmu$eV ($\sim$ 200 GHz),  is sufficient to realize a gatemon qubit \cite{2015_PRL_gatemon, Huo_gatemon}. In Fig. S7 \cite{SM}, we show a second Josephson device with tunable switching current (from 0 nA to 150 nA).

We then fix $V_{\text{G}}$ and scan $B$ along different axis, see Fig. 4(a) for the axis labeling. Figure 4(e) is the $B_z$ scan (aligned with the wire). The supercurrent is fully suppressed at $\sim$  1 T. As a contrast, the $B_y$ and $B_x$ scans in Figs. 4(f-g) reveal a much smaller critical field. Figure 4(h) plots line cuts along the three $B$ directions: At $|B|$ = 0.3 T, the supercurrent for the $z$-direction is clearly visible, while for the $y$ and $x$ directions, the supercurrent is suppressed. This anisotropic behavior of $B$ dependence is likely due to orbital effects. For $B$ aligned in the $z$ axis, i.e. parallel to the nanowire and In film, the orbital effects of $B$ is minimized. A higher critical field is thus expected. For the other two directions, $B$ is either perpendicular to the film or the nanowire. Orbital effects thus suppress superconductivity and a lower critical field is observed. The critical field of supercurrent is smaller than that of the gap, as supercurrent is generally more fragile and sensitive to $B$. For temperature dependence of the devices, see Fig. S8 \cite{SM}.   

\section{Conclusion}

In summary, we have realized In thin films on PbTe nanowires. Grain formation is prevented by low-temperature growth and in situ capping. The PbTe-In interface is atomically sharp, enabling the observation of a hard gap. The gap size of 1.18 meV is enhanced by a factor of two compared to the bulk In (0.5 meV). The $T_c$ of an In thin film on a PbTe substrate is $\sim$ 6 K, also significantly enhanced compared to the case of other substrates. Subgap states appear at finite $B$'s. The effective g-factors are significantly enhanced compared to bare PbTe nanowires, due to the presence of In. The gap can maintain large and hard for $T$ up to 1 K. In Josephson devices, supercurrents can be observed and gate-tuned. Our results on PbTe-In opens a new avenue for hybrid quantum devices. More importantly, the combination of In and PbTe enhances the properties of both: The large gap in In due to PbTe, and the enhanced g-factors in PbTe due to In. This hybridization effect is rare in previous superconductor-semiconductor hybrids, but desired for Majorana research.

\section{Acknowledgement} 

This work is supported by National Natural Science Foundation of China (92065206) and the Innovation Program for Quantum Science and Technology (2021ZD0302400). 

\section{Data Availability}

Raw data and processing codes within this paper are available \cite{rawdata}.

\bibliography{mybibfile}

\newpage

\onecolumngrid

\newpage


\section*{Supplementary material}

# Supplemental Material for "Epitaxial Indium on PbTe Nanowires for Quantum Devices"

Zuhan Geng,[1, *] Fangting Chen,[1, *] Yichun Gao,[1, *] Lining Yang,[1] Yuhao Wang,[1] Shuai Yang,[1] Shan Zhang,[1] Zonglin Li,[1] Wenyu Song,[1] Jiaye Xu,[1] Zehao Yu,[1] Ruidong Li,[1] Zhaoyu Wang,[1] Xiao Feng,[1, 2, 3, 4] Tiantian Wang,[2, 4] Yunyi Zang,[2, 4] Lin Li,[2] Runan Shang,[2, 4] Qi-Kun Xue,[1, 2, 3, 4, 5] Ke He,[1, 2, 3, 4, †] and Hao Zhang[1, 2, 3, ‡]

[1]*State Key Laboratory of Low Dimensional Quantum Physics, Department of Physics, Tsinghua University, Beijing 100084, China*
[2]*Beijing Academy of Quantum Information Sciences, Beijing 100193, China*
[3]*Frontier Science Center for Quantum Information, Beijing 100084, China*
[4]*Hefei National Laboratory, Hefei 230088, China*
[5]*Southern University of Science and Technology, Shenzhen 518055, China*

## Method

**Fabrication of shadow walls.** A negative resist, Hydrogen SilsesQuioxane (FOx-16, Dow Corning) was spun onto a CdTe(110) substrate at 2000 rpm and baked at 100 °C for 5 min. Patterns of shadow walls and markers were written using electron beam lithography (EBL) at a dose of 272 µC/cm$^2$. The chip was then developed in AR 300-44 for 150 s, then rinsed in DI water and IPA for 1 min and 30 s, respectively.

**Growth of Pb$_{0.97}$Eu$_{0.03}$Te buffer layer.** The CdTe(110) substrate was transferred to a molecular beam epitaxy (MBE) chamber. Argon bombardment (beam energy 2 keV, beam current 4.30 µA, 15 min) was performed in loadlock to remove surface oxide. The substrate was then annealed at 350 °C for 60 min. A 75-nm-thick Pb$_{0.93}$Eu$_{0.07}$Te film was grown at 285 °C. Finally, a 60-nm-thick CdTe capping layer was grown at room temperature.

**Fabrication of SiN mask**. A 36-nm-thick SiN mask was deposited on the substrate using plasma-enhanced chemical vapor deposition. Electron beam resist, AR 672.045, was spun at 4000 rpm, and baked at 150 °C for 5 min. Nanowire trench patterns along the [001] crystal direction were written by EBL at a dose of 380 µC/cm$^2$. The chip was developed in (MIBK: IPA = 1: 3) for 45 s and rinsed in IPA for 45 s. Reactive ion etching in a CHF$_3$/O$_2$ plasma (55 sccm CHF$_3$, 5 sccm O$_2$, 100 W, 75 s) was performed to etch the SiN mask. O$_2$ plasma (0.30 mbar, 30 sccm O2, 60 W, 30 s) was performed and the resist was dissolved in acetone for 60 min.

**PbTe nanowire growth.** The chip was loaded into an MBE loadlock followed by argon bombardment (beam energy 2 keV, beam current 4.30 µA, 5 min). The chip was then transferred into the MBE main chamber and was heated to 340 °C overnight to decompose the residue CdTe in the trenches. PbTe nanowires (thickness 150 nm) were selectively grown at 285 °C within the trenches on top of the Pb$_{0.97}$Eu$_{0.03}$Te buffer.

**Growth of In films.** The chip was transferred to a tilted sample stage in the interconnected MBE chamber without breaking the ultra-high vacuum. The substrate was cooled down to 90 ± 0.5 K for 60 min using liquid nitrogen. Indium was then grown (thickness 7-10 nm), with a tiled angle of 60 degrees for shadowing. Subsequently, a 10-nm-thick CdTe capping layer was grown. The sample stage was cooled by liquid nitrogen throughout the In and CdTe growth.

**Measurement.** Standard two-terminal measurement circuit was used for NS and SS devices. Contributions from the fridge filters have been subtracted from the voltage bias and conductance of the devices. For the Josephson SS device (Fig. 3), a contact resistance of 230 Ω was subtracted.

**STEM processing.** Due to the low melting point of In, the process of focused ion beam (FIB) was performed at low temperature (∼ -110 °C) using liquid nitrogen cooling. This can keep the morphology of the In film and prevent its melting.

---

* equal contribution
† kehe@tsinghua.edu.cn
‡ hzquantum@mail.tsinghua.edu.cn

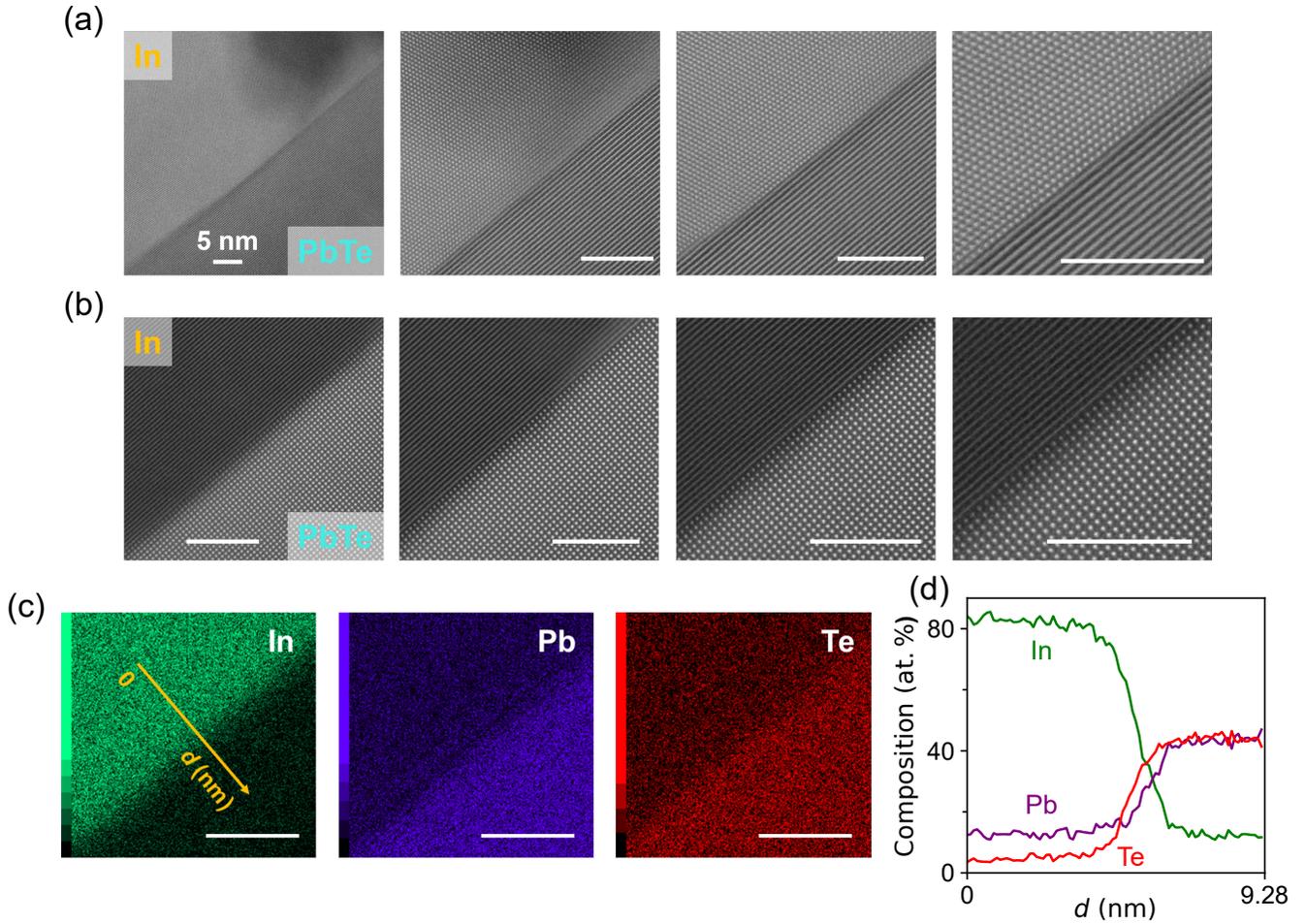

FIG. S1. (a-b) Additional STEM images of the PbTe-In interface along the [010] zone axis of In (a) and [001] zone axis of PbTe (b). (c) Energy-dispersive x-ray spectroscopy (EDX) maps of In, Pb, and Te, respectively. Scale bars in (a-c) are all 5 nm. (d) Composition ratio of each elements, along the orange arrow direction in (c).

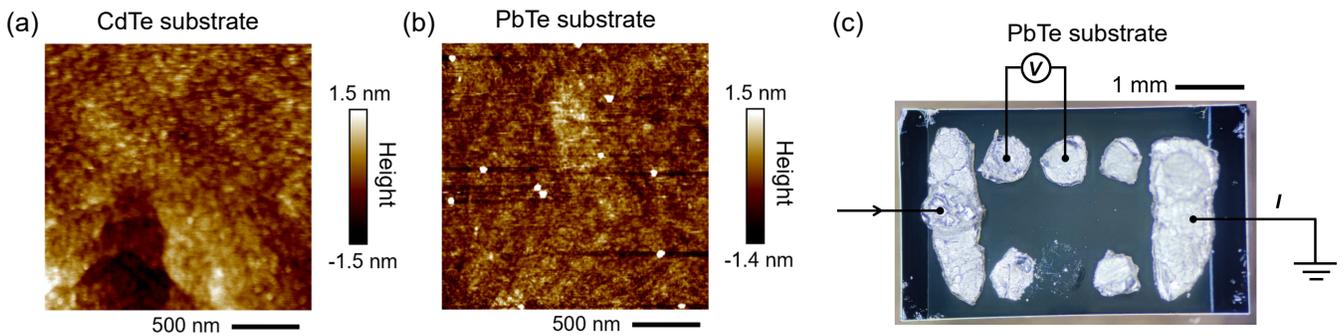

FIG. S2. (a-b) Atomic force microscope (AFM) images of the In films in Fig. 1(g), grown on a CdTe (a) and PbTe (b) substrate, respectively. The surface inhomogeneity is within ± 1.5 nm, much smaller than the film thickness (5 nm). The film was grown and capped by CdTe at low temperature using liquid nitrogen cooling. (c) Image of the In film and schematic of the four-terminal measurement circuit. The measurement was performed at 0 T.

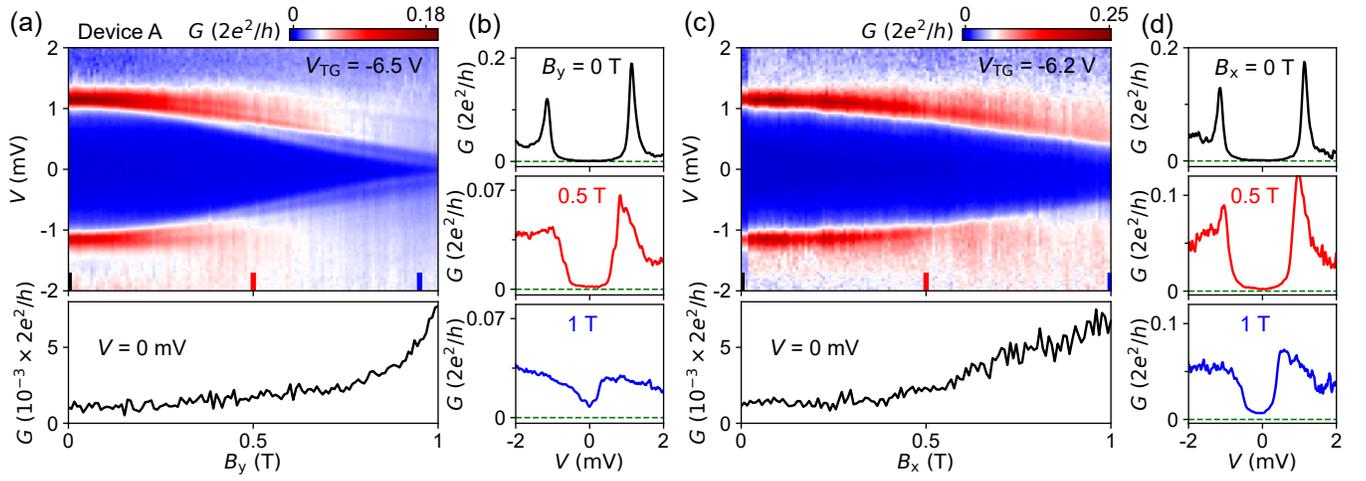

FIG. S3. $B$ scans along other directions for device A. (a) $B_y$ scan with zero-bias line cut shown in the lower panel. (b) Line cuts at 0 T, 0.5 T, and 1.0 T from (a). (c-d) $B_x$ scan.

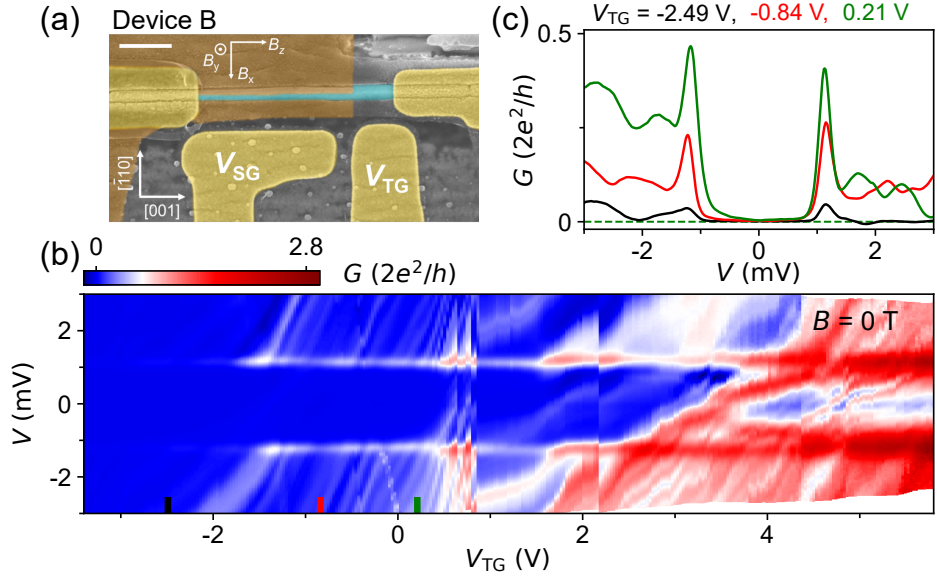

FIG. S4. (a) False-color SEM of device B. Scale bar, 500 nm. (b) $G$ vs $V$ and $V_{TG}$ of device B at $B = 0$ T. (c) Several line cuts from (b), see the corresponding color bars.

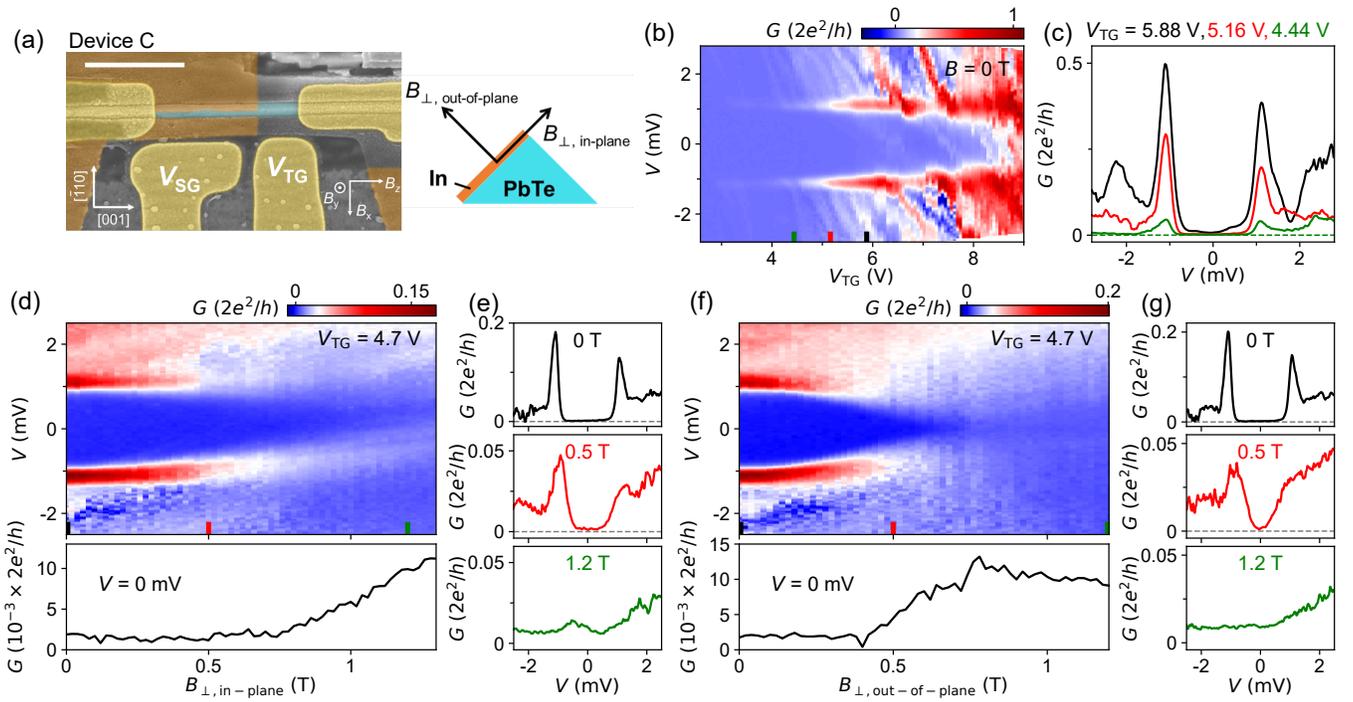

FIG. S5. (a) False-color SEM of device C (left) and the schematic of the cross-section (right). (b) $G$ vs $V$ and $V_{TG}$ of device C at $B = 0$ T. (c) Several line cuts from (b), see the corresponding color bars. (d) $B$ scan of the gap with $B$-direction being perpendicular to the PbTe nanowire but parallel to the In film, see the schematic in (a). Lower panel, zero-bias line cut. (e) Line cuts from (d). (f-g) $B$ perpendicular to both the In film and the PbTe nanowire.

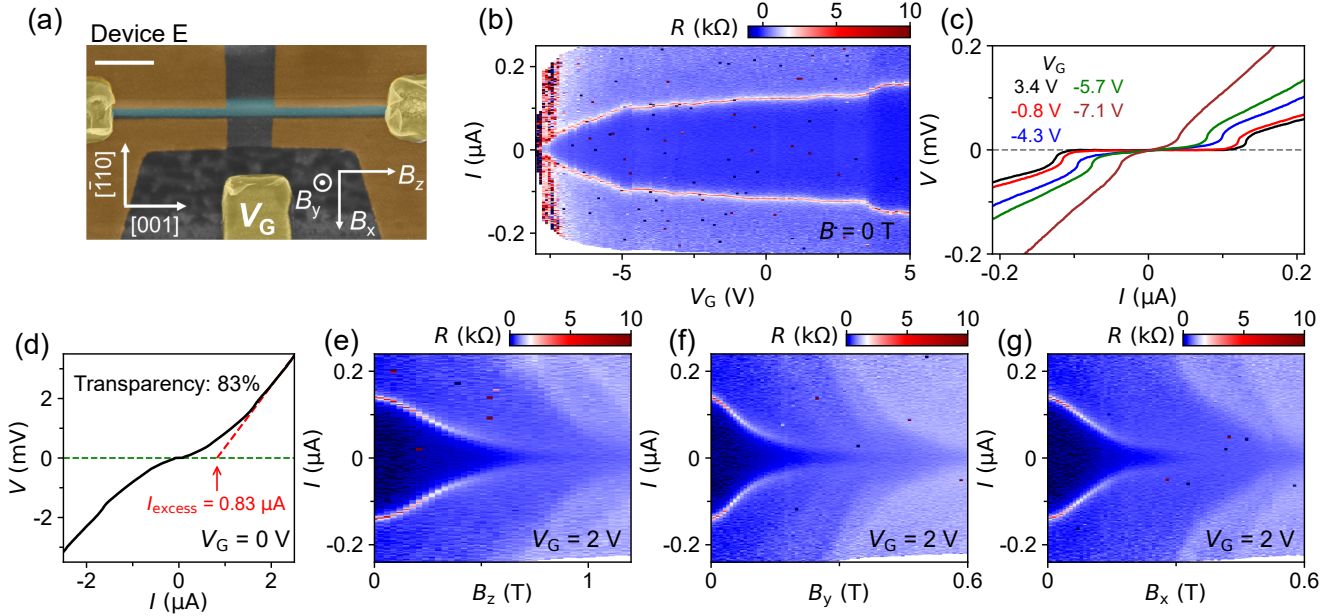

FIG. S6. A second Josephson device. (a) False-color SEM. Scale bar, 500 nm. (b) $R$ (numerical differentiation) vs $I$ and $V_G$. (c) Line cuts from (b). (d) $I$-$V$ over a larger $I$ range for the estimation of excess current and transparency, $\sim 83\%$. (e-g) $B$ dependence of the supercurrent along three axes (see labeling in (a)).



\end{document}